\documentclass[twocolumn]{aastex62}
\usepackage{amsmath,amstext}
\usepackage{comment}
\usepackage{apjfonts} 
\usepackage{chngcntr}
\usepackage{float}
\usepackage[overload]{textcase} 

\newcommand{\be}{\begin{eqnarray}}
\newcommand{\ee}{\end{eqnarray}}
\newcommand{\lp}{\left(}
\newcommand{\rp}{\right)}

\begin{document}

\normalsize
\interfootnotelinepenalty=10000
\title{Multi-Phase Shock Cooling Emission in Ultra-Stripped Supernovae}

\author{Annastasia Haynie}
\affiliation{Department of Physics and Astronomy, University of Southern California, Los Angeles, CA 90089, USA; ahaynie@usc.edu}
\affiliation{The Observatories of the Carnegie Institution for Science, 813 Santa Barbara St., Pasadena, CA 91101, USA}

\author{Samantha C. Wu}
\affiliation{The Observatories of the Carnegie Institution for Science, 813 Santa Barbara St., Pasadena, CA 91101, USA}
\affiliation{Center for Interdisciplinary Exploration \& Research in Astrophysics (CIERA), Physics \& Astronomy, Northwestern University, Evanston, IL 60202, USA}
\affiliation{California Institute of Technology, Astronomy Department, CA 91125, USA}

\author{Anthony L. Piro}
\affiliation{The Observatories of the Carnegie Institution for Science, 813 Santa Barbara St., Pasadena, CA 91101, USA}

\author{Jim Fuller}
\affiliation{TAPIR, Walter Burke Institute for Theoretical Physics, Mailcode 350-17, California Institute of Technology, Pasadena, CA 91125, USA}

\begin{abstract}

Ultra-stripped and Type~Ibn supernovae (USSNe and SNe Ibn, respectively) are fast-evolving, hydrogen-poor transients that often show signs of interaction with dense circumstellar material (CSM). \cite{Wu2022} identify a mass range for helium-core stars in which they expand significantly during core oxygen/neon burning, resulting in extreme late-stage mass loss in tight binaries ($P\sim1-100\,{\rm days}$). Here we explore the resulting light curves from a subset of models from \cite{Wu2022} and find that in some cases they can exhibit two phases of shock cooling emission (SCE). The first SCE is attributed to the circumbinary material, and the second SCE is from the extended helium-burning envelope of the exploding star. Since SCE luminosity is roughly proportional to the initial radius of the emitting material, events that exhibit both phases of SCE provide the exciting opportunity of measuring both the extent of the CSM and the radius of the exploding star. These light curves are explored with both analytic arguments and numerical modeling, and from this we identify the parameter space of CSM mass, helium envelope mass, and nickel mass, for which the helium envelope SCE will be visible. We provide a qualitative comparison of these models to two fast-evolving, helium-rich transients, SN2019kbj and SN2019dge. The similarity between these events and our models demonstrates that this extreme binary mass loss mechanism may explain some SNe~Ibn and USSNe.

\end{abstract}


\section{Introduction} \label{sec:intro}

Observations of core-collapse supernovae (SNe) during the first days to weeks after explosion can provide useful information about their progenitors \citep{Piro2013}. In particular, as the shock-heated material in a SN expands and cools, it produces a signature typically observed in the optical/UV called shock cooling emission (SCE; \citealp{Gradberg1976,Falk1977,Chevalier1992,nakar_sari2010,rabinak_waxman2011,Margalit2022a,Margalit2022b}). This can be an especially valuable tool because the SCE luminosity is roughly proportional to the initial radius of the emitting material \citep[e.g.,][]{nakar_sari2010}, which probes the mass distribution of the progenitor. In the last decade or so, SCE has thus been used to study the progenitors of a wide range of core-collapse SNe, including the radii of yellow supergiants that make Type Ib SNe \citep[e.g.,][]{Woosley1994,Bersten2012,Nakar2014} and probing the compact circumstellar material (CSM) around Type II SNe \citep[e.g.,][]{morozova2017,morozova2018,Jacobson2024}.

Recently, there has been a growing number of transients that show a fast rise and early bright blue emission that has been attributed to SCE from an extended helium-rich envelope \citep[e.g.,][]{De2018,Taddia2018,Ho2020,Jacobson2020,Yao2020,Pellegrino22a}. Narrow emission lines have been seen from some of these events, which are designated SN Ibn, and when the ejected masses are inferred to be especially low ($\lesssim1\,M_\odot$) they are often referred to as ultra-stripped SNe (USSNe). In some cases, even helium-poor Type Icn SNe show signatures that suggest they may be from a similar channel to the USSNe \citep{Pellegrino2022b}. The origin and fate of these systems are of much interest since in some cases they may be the progenitors of compact neutron star binaries \citep{Dewi2003,Tauris2013,Tauris2015}.

In previous work, ``case BB'' mass transfer has been invoked to produce these binaries with significant mass stripping \citep{Yoon2010,Tauris2013,Tauris2015,Zapartas2017,Laplace2020,ercolino2024}. Although these models could often replicate the low ejecta masses observed, most models do not predict the large amount of CSM needed to produce the early bright emission seen from many USSNe and Type Ibn SNe. However, many of the stripped progenitor models omit the evolution onward from oxygen/neon (O/Ne) burning, and thus miss crucial physics from these final years of the star’s lifetime. \cite{Wu2022} identified a range of helium-core masses ($\approx 2.5 - 3 M_{\odot}$) within which the stellar envelope expands significantly due to He-shell burning while the core burns O/Ne. This induces rapid late-stage mass transfer when evolved with a binary companion. They found that models with longer orbital periods tend to have mass loss rates that increase significantly in the {\it months to years} before silicon burning, leading to extremely late stage mass loss. In contrast, models with shorter orbital periods see a rise in mass loss rates {\it years to decades} before silicon burning, so the resulting CSM properties can vary greatly. \cite{Wu2022} hypothesized that the diverse properties of the CSM inferred from the SCE of USSNe and SNe Ibn could be related to these processes.

In this work, we follow up on this prediction by \cite{Wu2022} by exploring the light curves from these progenitors both analytically and numerically. We find that these models can sometimes result in two phases of SCE, with the first caused by the dense circumbinary material (which we hereafter refer to as the CSM) and the second by the extended helium envelope (HE). Since the SCE brightness is roughly proportional to the initial radius of the emitting material, events that show both phases of SCE offer the possibility of providing multiple probes into measuring critical radii within the mass distribution of these systems. Whether or not both phases of SCE are observable depends on a number of factors, including explosion energy, CSM mass, HE mass and nickel mass, and we explore when the SCE from the HE should be seen.

We begin in Section~\ref{sec:analytic} by examining the density profiles of the models from \cite{Wu2022} at the onset of core-collapse and then use analytic arguments to anticipate what the SCE luminosity should be from the different density components. We follow this by numerically exploding these models and computing their light curves. We discuss the methods we employ for this in Section~\ref{sec:methods} and summarize our results in Section~\ref{sec:sce}. In Section~\ref{sec:UUSNe}, we provide a qualitative comparison to two well studied events with helium-rich envelopes, SNe 2019dge \citep{Yao2020} and 2019kbj \citep{Ben-Ami2023}, which is followed by a summary and conclusion in Section~\ref{sec:conclusion}.


\section{Analytic Motivations}\label{sec:analytic}

\begin{figure}
\includegraphics[width=0.48\textwidth]{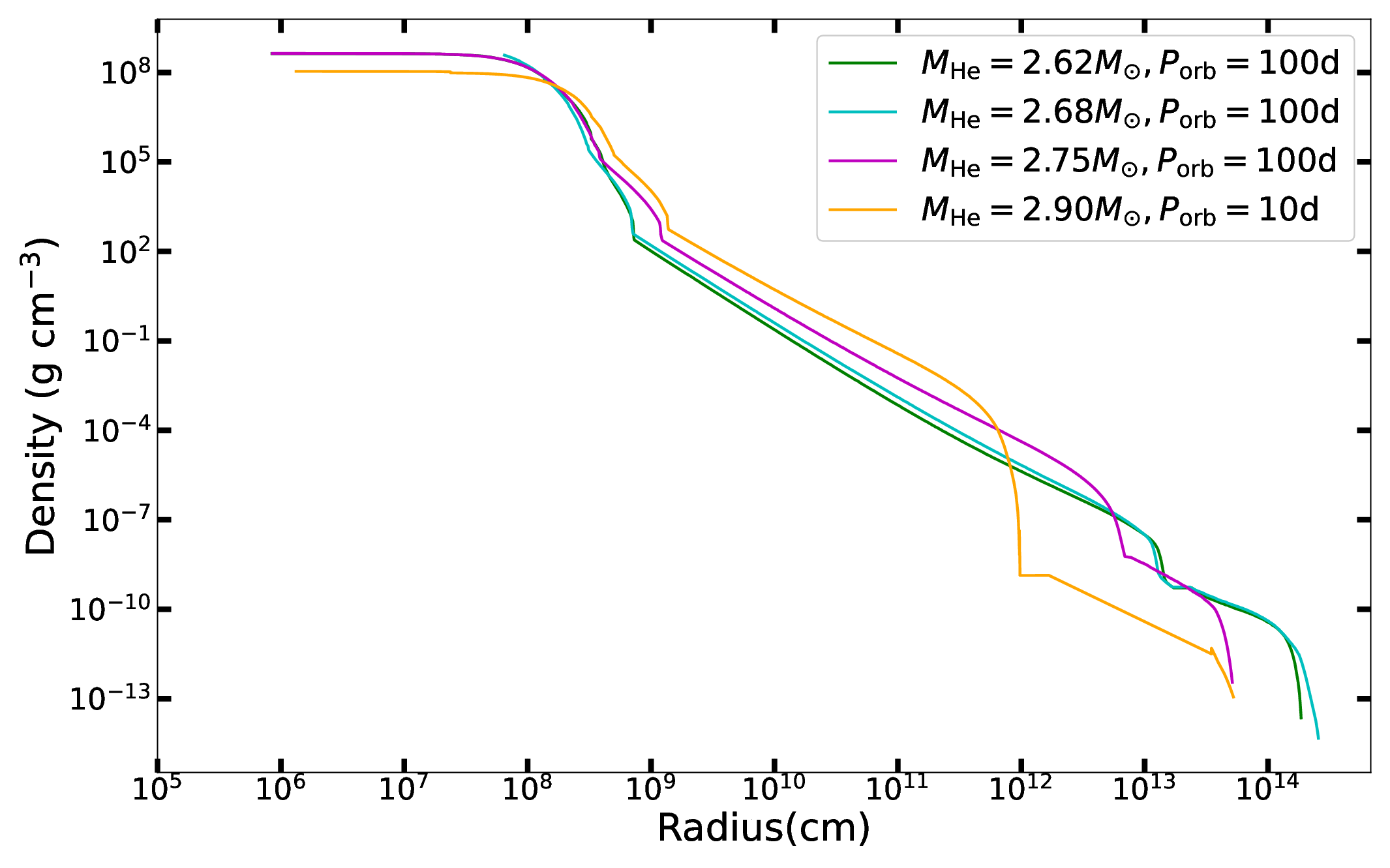}
\caption{Density profiles of a subset of the models from \cite{Wu2022} at the onset of core-collapse, prior to the excision of the inner neutron star. We see that each profile can be broken down into three distinct regions, (1) an inner dense core ($R \lesssim 10^9$cm), (2) an extended HE ($10^9 \lesssim R \lesssim 10^{13}$cm) , and (3) a helium-rich CSM above. The labels correspond to the initial helium core mass and orbital period of each model.}
\label{fig:profiles}
\end{figure}

In Figure~\ref{fig:profiles}, we show the density profiles of a subset of the \cite{Wu2022} models and see that they can be roughly broken into three sections: (1) an inner core up to $\sim 10^9\,$cm, (2) a HE that extends out to $\sim10^{12}-10^{13}\,{\rm cm}$ depending on the model, and (3) an outer, low density CSM. The shock that unbinds the star will generate radioactive nickel in the inner layers and then subsequently heat the stellar material as it passes through the star. The exploding star will then expand and cool, producing the SCE that powers the light curve \citep[e.g.,][]{nakar_sari2010}.

We next use these density profiles to estimate the properties of this SCE. For this we follow analytic arguments presented in \cite{piro2021}, although we note that similar results are available in other studies \citep[e.g.,][]{nakar_sari2010,rabinak_waxman2011,Margalit2022b}. One advantage of \cite{piro2021} is that it is specifically focused on emission from extended material with a mass that is lower than the core mass, as well as being calibrated to match numerical simulations using the SuperNova Explosion Code \citep[\texttt{SNEC},][which we describe further in Section~\ref{sec:methods}]{SNEC}. This work finds that the SCE bolometric luminosity decreases as a power law initially up until the diffusion time, $t_d$, which we rewrite as
\be
    t_d \approx 2.7\lp \frac{\kappa M_{\rm ext}^{3/2}}{E_{\rm ext}^{1/2}c} \rp ^{1/2}, \label{eq:td}
\ee
where $M_{\rm ext}$ is the mass of the extended material producing SCE and $E_{\rm ext}$ is the energy imparted on the extended material by the shock. After $t_d$, the bolometric luminosity declines exponentially (for a more detailed discussion of diffusion time and the derivation of the following scaling, see Section 2 of \citealp{piro2021}). At the break between these two scalings, the luminosity is 
\be
    L_{\rm SCE}\lp t_d \rp
    \approx 0.11 \frac{c R_{\rm ext} E_{\rm SN}}{\kappa M_{\rm ext}}
    \lp \frac{M_{\rm ej}}{3 M_{\odot}} \rp ^{-0.7}
    \lp \frac{M_{\rm ext}}{0.01 M_{\odot}} \rp ^{0.7}, \label{eq:Lum}
\ee
where $c$ is the speed of light, $R_{\rm ext}$ is the radial extent of the extended material, $\kappa$ is the specific opacity, and $M_{\rm ej}$ is the total mass of the ejecta underneath the CSM, including both the HE and inner core\footnote{\cite{piro2021} uses the notation $M_{\rm c}$ to refer to the underlying core material. We choose to relabel this parameter to provide distinction between the helium envelope and inner core.}. 

We use this framework to estimate the time-dependent luminosity during each of the phases of SCE from the two regions, $L_{\rm SCE}^{\rm CSM}(t)$ and $L_{\rm SCE}^{\rm HE}(t)$, for the CSM and HE, respectively. When calculating $E_{\rm ext}$ for the CSM, we use Equation~(23) from \cite{piro2021}, which estimates how much energy is transferred into the CSM by the shock. For the HE, its mass exceeds the mass of the underlying core and thus we take $E_{\rm ext}$ for the HE to be the entire explosion energy. For this reason, $L_{\rm SCE}^{\rm HE}$ should be viewed as an upper limit.

\begin{figure}
\includegraphics[width=0.47\textwidth]{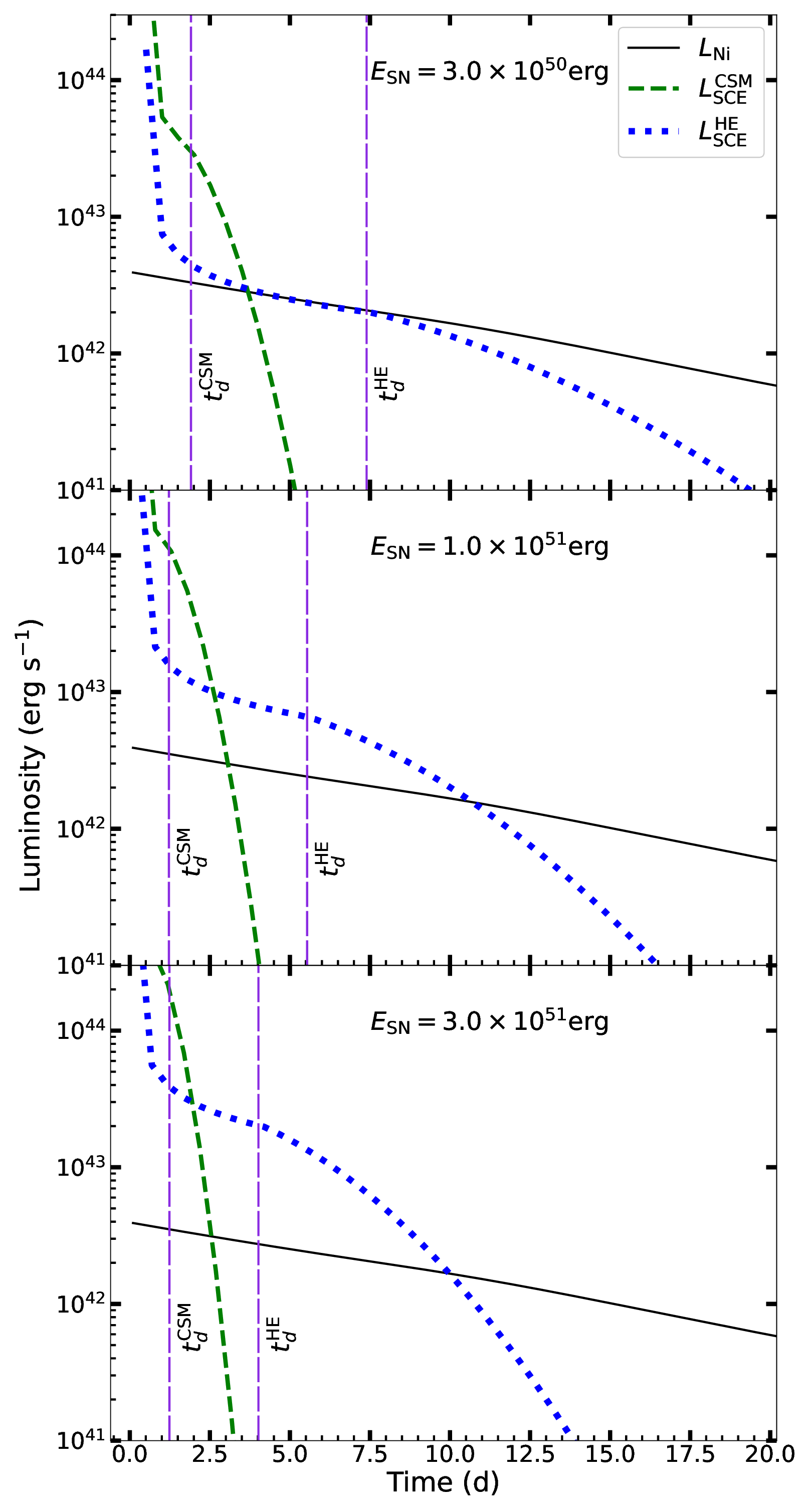}
\caption{The analytic light curves of a fiducial model at three different explosion energies as labeled. Each light curve has contributions from three different components, (1) SCE from the CSM, $L_{\rm SCE}^{\rm CSM}$ (green-dashed line), (2) SCE from the HE, $L_{\rm SCE}^{\rm HE}$ (blue-dotted line), and (3) instantaneous nickel heating, $L_{\rm Ni}$ (black-solid line).  See the text for full discussion of how these components are calculated. Vertical solid gray lines denote the diffusion time for both the CSM and HE. At the lowest energy (top panel), the luminosity due to nickel heating slightly outshines the HE SCE at $t_{\rm d}^{\rm HE}$, but as $E_{\rm SN}$ increases, the HE SCE becomes bright enough to be a distinct feature in the overall light curve.}
\label{fig:analytic}
\end{figure}

The resulting analytic luminosities are presented in Figure~\ref{fig:analytic} using the $M_{\rm He}=2.75\,M_\odot$, $P_{\rm orb}=100\,{\rm d}$ progenitor from \cite{Wu2022} as a fiducial model  (the purple line in Figure~\ref{fig:profiles}). $L_{\rm SCE}^{\rm CSM}(t)$ and $L_{\rm SCE}^{\rm HE}(t)$ are plotted as green-dashed and blue-dotted lines, respectively. For these we set $\kappa=0.2\,{\rm cm^2\,g^{-1}}$ as is appropriate for helium-rich material. We also plot the instantaneous luminosity due to the radioactive decay of $^{56}{\rm Ni}$ ($L_{\rm Ni}$, assuming $M_{\rm Ni} = 0.05 M_{\odot}$) as a solid-black line for which we include the partial leakage of gamma-rays according to \cite{Haynie2023} using a fiducial gamma-ray escape time of $T_0 = 30\,$days. The three panels use successively higher explosion energies, $E_{\rm SN}$, from top to bottom as labeled. In each case, $L_{\rm SCE}^{\rm CSM}( t_d^{\rm CSM})$ is brighter than $L_{\rm SCE}^{\rm HE}(t_d^{\rm HE})$ due to having a larger radius. Conversely, the mass of the HE is larger than the mass of the CSM, and will therefore have a characteristically longer diffusion time.

By varying the explosion energy, we see that $L_{\rm SCE}^{\rm CSM}$ becomes brighter for higher $E_{\rm SN}$ but also declines more quickly as $t_d^{CSM}$ becomes shorter. The SCE from the CSM is always prominent due to its especially large radius. $L_{\rm SCE}^{\rm HE}$ behaves similarly, becoming brighter and shorter lived with larger $E_{\rm SN}$. For the lowest explosion energy, we find $L_{\rm SC}^{\rm HE}<L_{\rm Ni}$, while as explosion energy increases $L_{\rm SC}^{\rm HE}$ becomes more noticeable and eventually $L_{\rm SC}^{\rm HE}>L_{\rm Ni}$ so that HE SCE can be observed directly.

This comparison opens up the possibility that due to the unique density structure found in the models of \cite{Wu2022}, the light curve features can probe multiple distinct regions within the interior of the progenitors. It also shows that whether this is possible depends on the combination of $M_{\rm CSM}$, $M_{\rm HE}$, $M_{\rm Ni}$, and $E_{\rm SN}$. Motivated by this, we next turn to more detailed numerical modeling to understand this parameter space and the diversity of light curves.

\section{Models and Methods}\label{sec:methods}

We use the open-source, one-dimensional, hydrodynamic radiative transfer code SNEC \citep{SNEC} to generate light curves for each of the models shown in Figure~\ref{fig:profiles}. As is described in \cite{morozova2018}, prior to explosion we assume that the inner core of each model forms a neutron star and excise it at the Si/O interface, using a ``boxcar'' method with a $0.4M_{\odot}$ width in mass space to smooth out the remaining profile over 4 iterations. We set the total nickel mass to be $M_{\rm Ni} = 0.05M_{\odot}$ and use the moderate nickel mixing scheme as described in \cite{Haynie2023}, such that radioactive $^{56} \mathrm{Ni}$ is mixed through the inner 50\% of the ejecta in mass space. Because treatments of nickel mixing are not well agreed upon in CCSNe modeling, we use this moderate scheme as a baseline and future explorations of these progenitors may find it important to explore the impacts of varying nickel mixing schemes on USSNe and Type Ibn light curves. 

\begin{figure}
\includegraphics[width=0.48\textwidth]{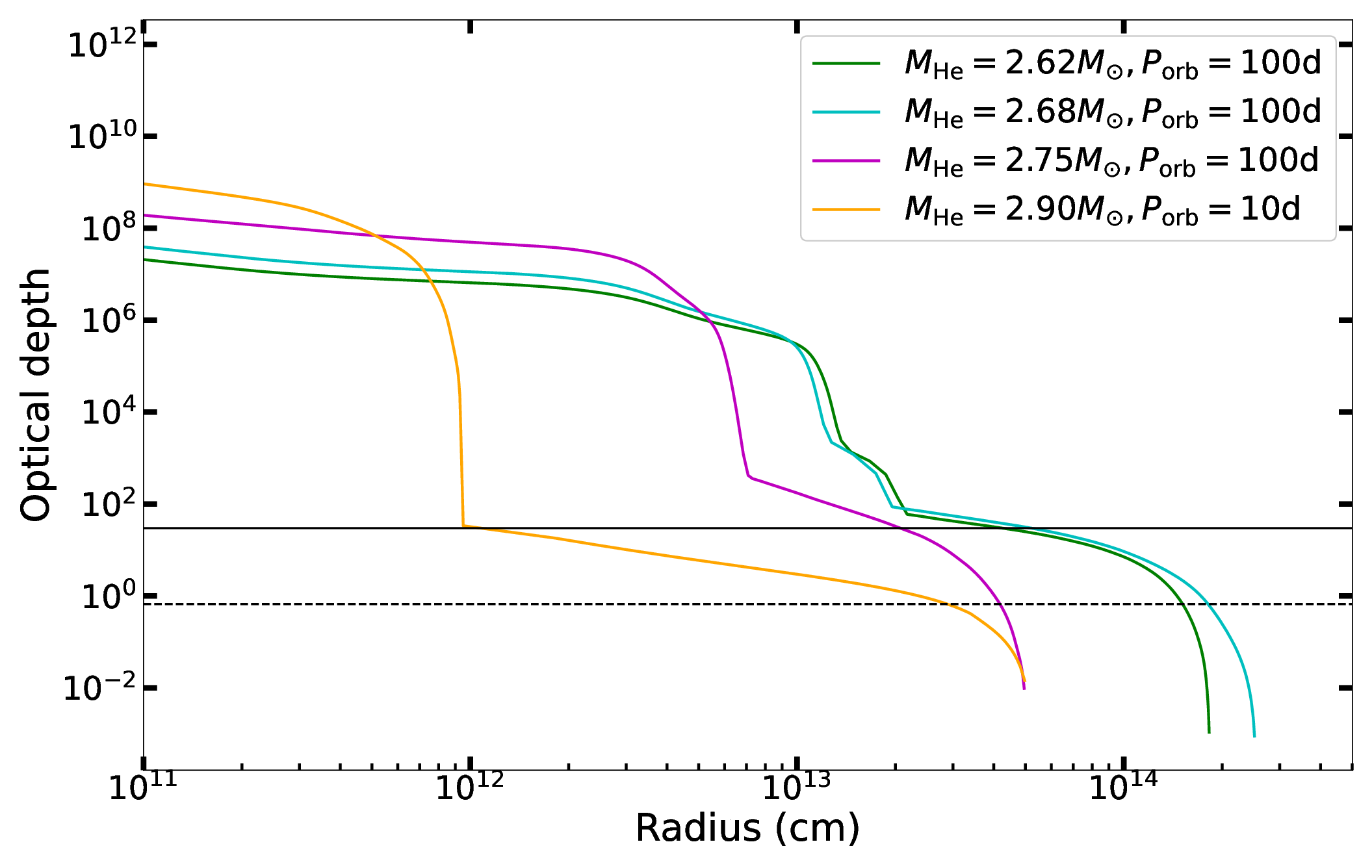}
\caption{Optical depth profiles at time $t = 0$ as calculated by \texttt{SNEC} using OPAL Opacity tables of each model from Figure~\ref{fig:profiles} (colored line correspond to the same models). Solid and dashed lines represent $\tau \approx c/v\approx30$ and $\tau=2/3$, respectively. These roughly correspond to the depth of the luminosity shell and photosphere, both of which must be resolved in the light curve modeling. Many models have some amount of optically thin material above the photosphere that is not contributing to the light curve, thus we  remove material above $\tau=2/3$ to prevent numerical difficulties in our simulations.
}
\label{fig:tau}
\end{figure}

Finally, we set a floor value for the opacity to further prevent numerical errors due to tracking extremely diffuse material and to effectively control the amount of recombination that can occur. Here, we choose an opacity floor of $0.001 {\rm cm}^2{\rm g}^{-1}$, roughly 100 times smaller than the electron scattering value for singly ionized helium, to allow for recombination without any constraint. An additional complication is that several of the models have especially extended and diffuse CSM due to the mass loss process explored in \cite{Wu2022}. Figure~\ref{fig:tau} shows the initial optical depth profiles, which demonstrates which regions have $\tau<2/3$. We elect to remove the outermost material above the photosphere to avoid numerical issues in our modeling. Otherwise, the low density material is being accelerated to high speeds by the shock and expanding rapidly. Since the material is optically thin, this removes very little mass from the progenitor and only minimally impacts the SCE.

\section{Results} \label{sec:sce}

\begin{figure}
\includegraphics[width=0.48\textwidth]{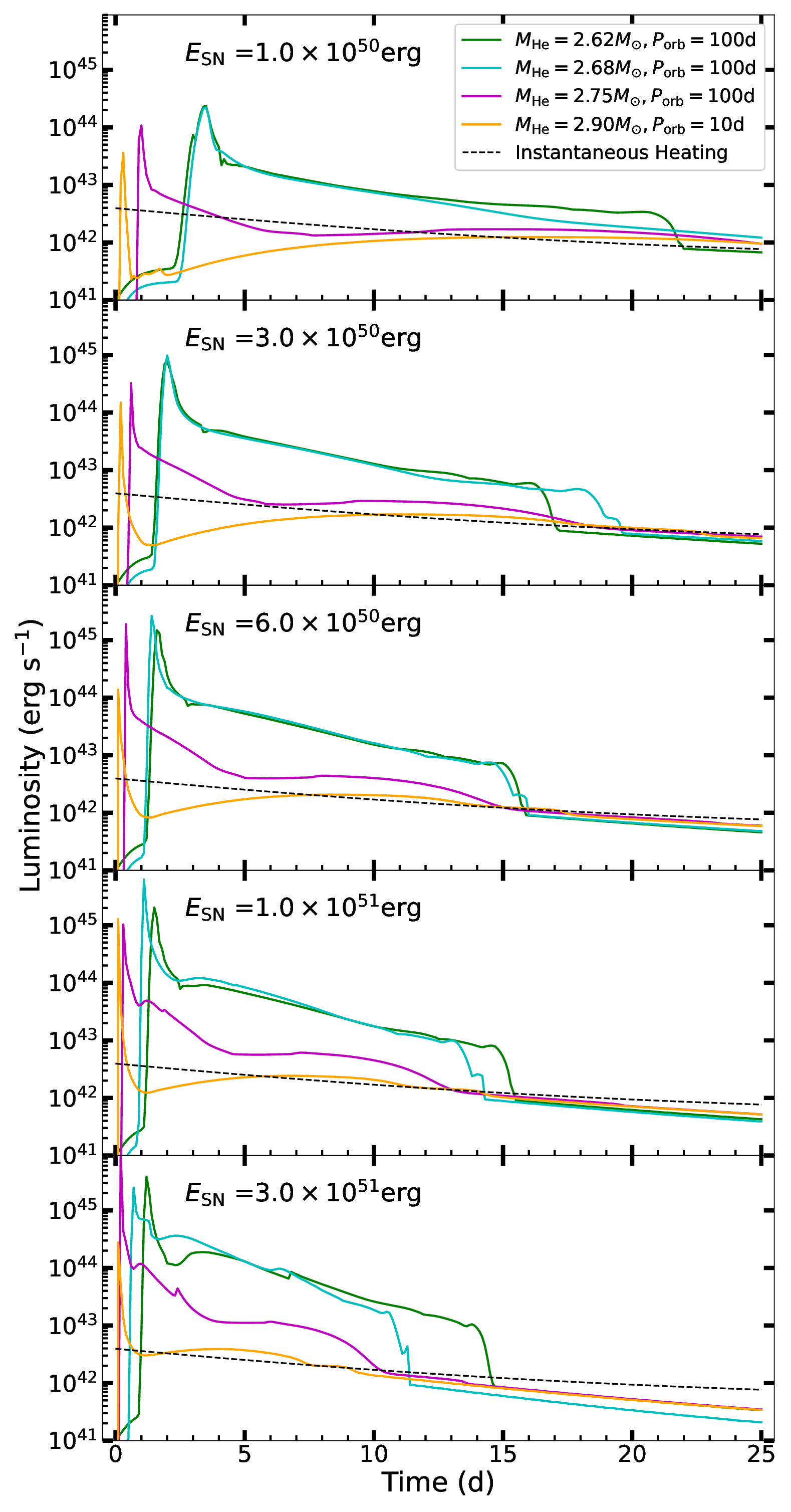}
\caption{Light curves (solid lines) for each stellar model at each explosion energy compared to the instantaneous $^{56}$Ni heating curve not including gamma-ray leakage (black dashed). As the energy increases, we see both increased structure in the early light curve and a greater excess in luminosity beyond the initial shock cooling phase. }
\label{fig:lightcurves1}
\end{figure}

The finalized models are exploded using a ``thermal bomb'' mechanism with five different injected energies from $E_{\rm SN} = 10^{50}\,{\rm erg}$ to $3\times 10^{51}\,{\rm erg}$, which is motivated by the range of values found by \cite{Lyman2016}. Because these are quickly evolving events, we model each SN out to 25 days beyond the time of explosion. By this time, all of the models are dominated by heating from radioactive nickel and we expect the rest of the light curve to resemble that of a nickel-powered CSM-free SN. All of the bolometric light curves from the simulations we ran are summarized in Figure~\ref{fig:lightcurves1}.

As we expect from the work by \citet[][also see \citealp{Chevalier2011}]{Haynie2021}, light curves of models that have denser, more massive CSM have peak luminosities due to shock breakout that are brighter and last longer than models with more diffuse CSM because the shock must travel further into the CSM to reach the optical depth threshold for the shock to breakout. We note that for the $M_{\rm He} = 2.90 M_{\odot}$ model, due to the small radius and very low CSM mass, the shock breakout width at the two highest explosion energies is shorter than the time resolution of this simulation. The peak luminosity of the signal may therefore be somewhat brighter than what is shown in the final two panels of Figure~\ref{fig:lightcurves1}, however this does not qualitatively change the shape of the light curve, especially during the later SCE phases that we are most interested in.

The light curves have a fair amount of structure at early times that becomes more pronounced as explosion energy increases. Furthermore, we see that in some models there is an enhanced luminosity following the initial SCE. We also plot the instantaneous heating rate from $^{56}$Ni (shown as the black dashed curve in Figure~\ref{fig:lightcurves1}). If these events simply followed ``Arnett's rule'' \citep{arnett1982}, then the peak bolometric luminosity would intersect the $^{56}$Ni curve as is seen for typical Type Ibc SNe. Instead, beyond the diffusion time relevant for CSM SCE, many models are much brighter than what is expected from only radioactive decay, indicating an additional source of energy.

\begin{figure}
\includegraphics[width=0.48\textwidth]{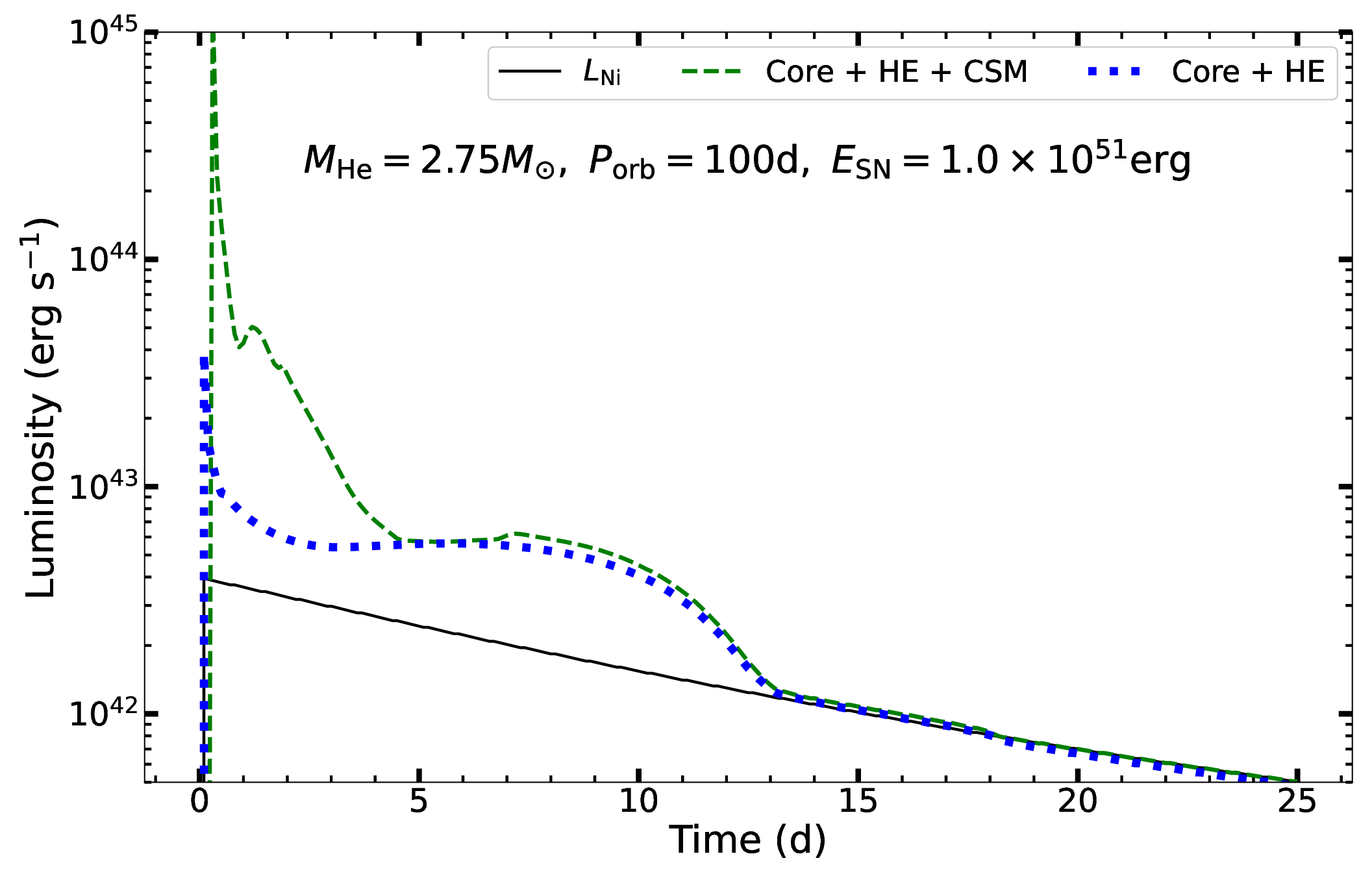}
\caption{Two light curves of a fiducial model with (green-dashed line) and without (blue-dotted line) the CSM. The black curve is the instantaneous $^{56}$Ni heating rate without including gamma-ray leakage. Initially the model with CSM is much brighter, due to the associated SCE. The models begin to coincide at $\sim5\,$days beyond the time of explosion. Thus at this point the CSM is no longer contributing to the light curve, and the luminosity excess from $\sim5-13\,$days must be due to SCE from the HE.} 
\label{fig:nocsm}
\end{figure}

To demonstrate that the excess luminosity is indeed due to SCE from the HE, we compare the light curves of a fiducial model, $M_{\rm He\ Core} = 2.75 M_{\odot}, E_{\rm SN} = 10^{51}\,{\rm erg}$, with and without CSM, shown in Figure~\ref{fig:nocsm}. When we remove the CSM by just chopping off this material, it changes the overall mass of the system by less than $\sim 5\%$, yet it reduces the total radius by $\sim 85\%$. This is reflected in the drastic reduction in the initial SCE over the first 5 days, after which the models coincide and are both much brighter than expected from radioactive decay alone until they decline into their radioactive decay tail at $\sim$13 days beyond the time of explosion. This shows that indeed the hump between $\sim5-12\,{\rm days}$ is from HE SCE.

\begin{figure}
\includegraphics[width=0.48\textwidth]{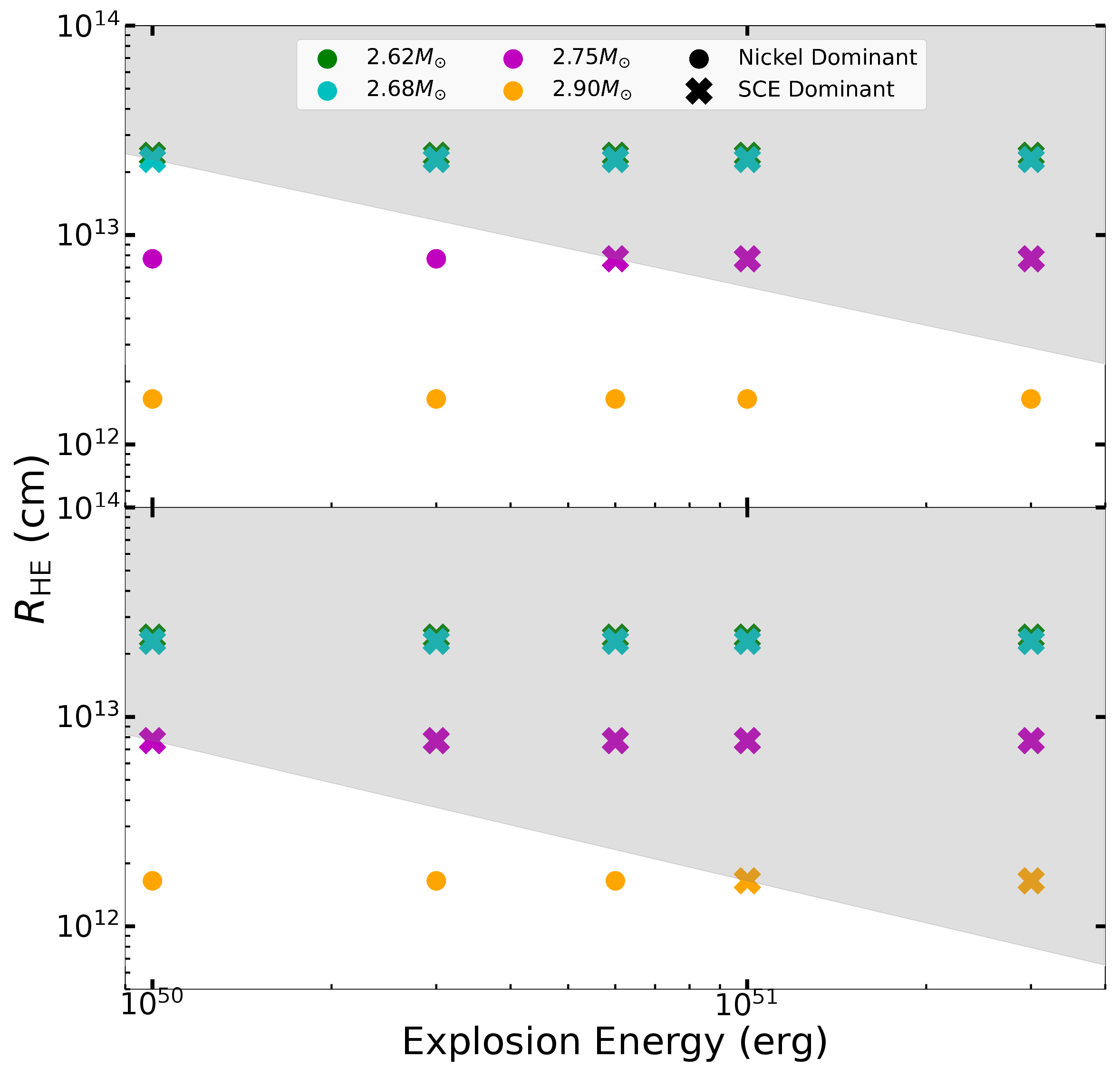}
\caption{Here we show the parameter space covered by our models exploded with $M_{\rm Ni} = 0.05 M_{\odot}$ (top) and $M_{\rm Ni} = 0.01 M_{\odot}$ (bottom). Points denoted by crosses have HE SCE dominant over the nickel luminosity, while those donated with circles are the opposite. This shows that HE SCE is stronger at higher explosion energies, and we have added a shaded region to guide the eye. The decrease in nickel mass in the bottom panel results in a decrease in the nickel luminosity by a factor of 5, requiring less energy for HE SCE to dominate, and we see the shaded region move to the right accordingly.}
\label{fig:ratio}
\end{figure}

In the above fiducial model it is easy to observe the impacts of the HE SCE as energy increases. However, for our two lowest helium-core mass models (the blue and green curves in Figure~\ref{fig:lightcurves1}), there is a significant excess of luminosity for all explosion energies, even though the HE SCE only begins to create a distinct bump in the light curve at $E_{\rm SN} = 3 \times 10^{51}$erg. This is because the mass of the CSM in these models is actually comparable to the mass of the HE, giving the two regions similar photon diffusion timescales. We conclude that the CSM SCE always dominates due its larger radius, but the combination of SCE from both regions creates a bright plateau before sharply declining into the decay tail once the helium has recombined. It is not that there is no HE SCE in these models, but rather the SCE from the CSM and HE are merging together.

While our survey of models shows that the relative sizes of the CSM and HE masses plays an important role in determining the light curves through the two diffusion timescales, the SCE luminosity is driven mainly by radius and explosion energy. In addition to the two lowest helium-core mass models having comparable CSM and HE diffusion timescales, these models also have the largest HE radius, further aiding their bright plateau. The relative masses seem to influence the light curve shape (bump-like or plateau-like), whereas the HE radius determines whether or not the HE SCE outshines the nickel decay. Motivated by this, in Figure~\ref{fig:ratio} we plot HE radius, $R_{\rm HE}$, versus the explosion energy, and for each model we consider whether HE SCE or $L_{\rm Ni}$ is dominant as denoted by the different symbols. We consider the HE SCE to be dominant when $L_{Ni}$ contributes less than half of the total luminosity on the timescale $t_d^{\rm HE}$. From this we see in Figure~\ref{fig:ratio} that there is a critical energy dividing these regimes, which we highlight with gray shading. We conclude that for higher explosion energies we  expect HE SCE to be more prominent.

We also explore how the relative size of the HE SCE and $^{56}$Ni luminosities change when we decrease the $^{56}$Ni mass. As long as the peak timescale does not change by too much, we know from Arnett's Rule that the $^{56}$Ni luminosity is roughly proportional to $M_{\rm Ni}$, so naturally, if $M_{\rm Ni}$ decreases, less energy will be required for the HE SCE to dominate. This is indeed seen in the bottom panel of Figure~\ref{fig:ratio}, where all models were exploded with $M_{\rm Ni} = 0.01 M_{\odot}$. We expect a factor of 5 decrease in the explosion energy needed for HE SCE to overpower the nickel luminosity compared to models in the top panel, and accordingly see the line dividing the two regimes move to the left in the bottom panel relative to the top panel.

\section{Comparison with SNe Ibn and USSNe} \label{sec:UUSNe}

As discussed in Section~\ref{sec:intro}, many USSNe and SNe~Ibn show a combination of bright early emission (suggesting SCE from especially extended material) and fast light curve evolution (indicating a relatively small ejecta mass). This motivated \cite{Wu2022} to suggest that the ultra-stripped progenitors they were investigating could be related to these events. To better test this, we next qualitatively compare our results to two well studied, fast-evolving, helium-rich transients, SN2019kbj and SN2019dge, which are classified as a SN~Ibn and USSN, respectively.

\subsection{SN 2019kbj}

SN2019kbj was discovered by the Asteroid Terrestrial-impact Last Alert System \citep[ATLAS,][]{Tonry2018, Tonry2019, Smith2020} on 2019 July 1 and was originally classified as a SN~II \citep{Hiramatsu2019} but was later reclassified as SN~Ibn due to the presence of narrow HeI emission lines in spectra taken one week after discovery \citep{Arcavi2022}. It is photometrically similar to other SNe~Ibn, including in post-peak decline rate (see \cite{Ben-Ami2023} Figure 2), but is was thought to require an additional power source in conjunction with radioactive decay to be able to reproduce the light curve. \cite{Ben-Ami2023} argued that this required interaction with a uniform density shell of CSM, rather than a steady-state wind that is commonly inferred for SNe~Ibn. This is interesting because it potentially suggests diversity in SNe~Ibn progenitors (also see \citealp{Hosseinzadeh2019}) and results in lower estimates of both $M_{\rm ej}$ and $M_{\rm Ni}$ ($\sim0.2-1.4\,M_{\odot}$ and $\sim0.08-0.1\,M_{\odot}$ respectively).

\begin{figure}
\includegraphics[width=0.48\textwidth]{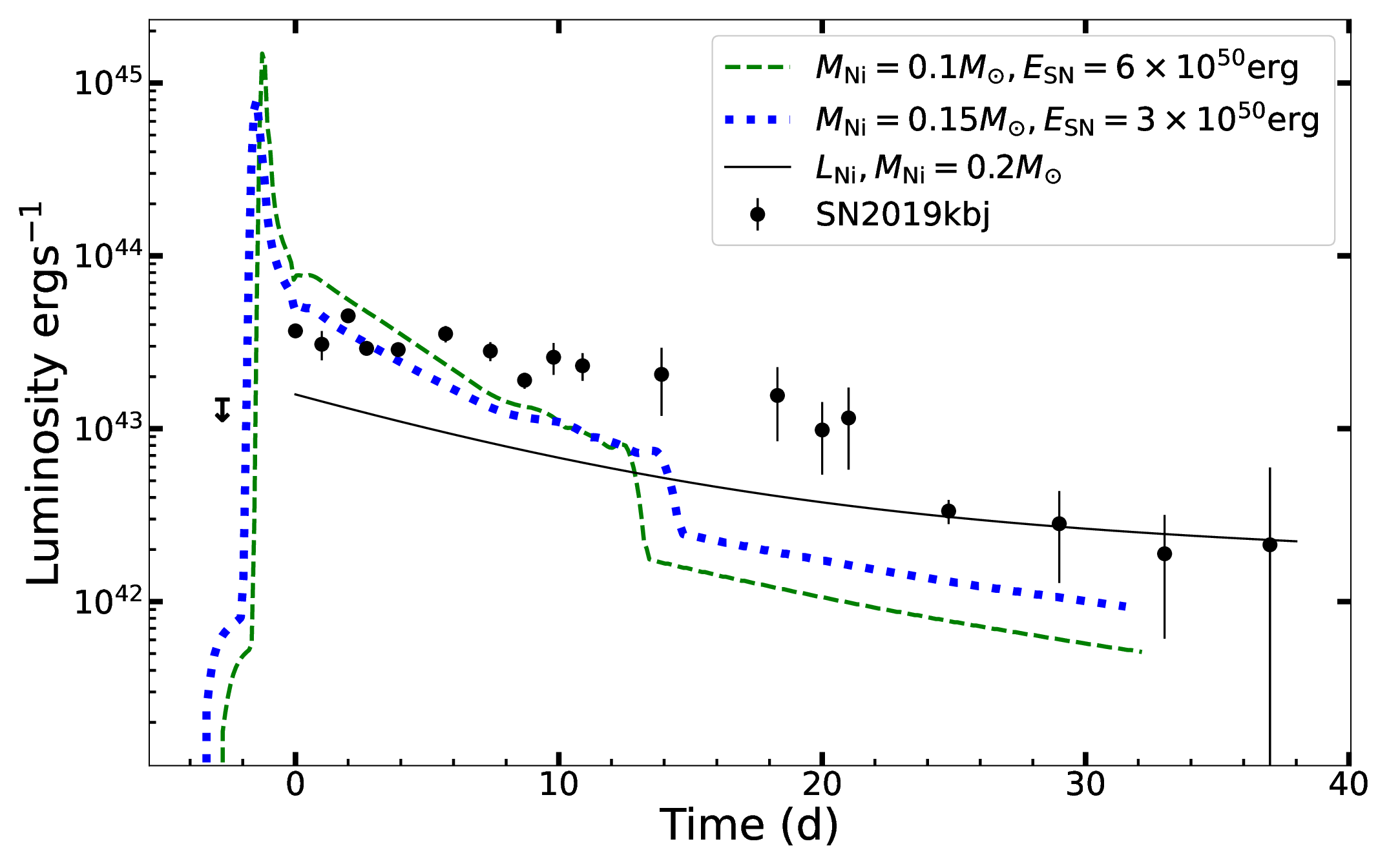}
\caption{Bolometric light curve of SN2019kbj (black dots) compared to two $M_{\rm He\ core} = 2.62 M_{\odot}$ models exploded with different combinations of explosion energy and nickel mass. We are not able to recreate the full length of the plateau out to $\sim$ 20 days, implying that the true progenitor has a higher ejecta mass than our models, which is in line with the estimates given in \cite{Ben-Ami2023}.}
\label{fig:kbj}
\end{figure}

Figure~\ref{fig:kbj} shows the inferred bolometric light curve of SN2019kbj \citep{Hosseinzadeh2020}. The presence of a long plateau is reminiscent of the morphology of the light curves presented in Figure \ref{fig:lightcurves1}, motivating us to compare our models to this event. We compare this with two light curves of $M_{\rm He\ Core} = 2.62 M_{\odot}$ at varying explosion energies and nickel masses. Our models favor a higher nickel mass to match the decay tail beyond $\sim$21 days, which agrees with the conclusions of \cite{Ben-Ami2023} for the steady-state wind scenario. Yet our model does not fully recreate the long plateau seen in the first 20 days of SN2019kbj, suggesting that our HE mass, which is on the lower end of the range suggested by the authors, is too small. Still, similarities in the qualitative appearance of the model and observed light curves suggest that stripped helium stars, perhaps with larger helium envelope masses at core collapse, may be good candidates for the progenitor of this event.

\subsection{SN 2019dge}

SN2019dge is a helium-rich ultra-stripped supernova first observed by Zwicky Transient Facility \citep[ZTF,][]{Belm2019, Graham2019} on 2019 April 7 and is characterized by a fast rise ($\lesssim 3\,{\rm days}$) but a longer decay time and fainter peak r-band magnitude ($-16.3\,{\rm mag}$) than typical SNe~Ibc \citep{Yao2020}. Early-time and follow-up spectra indicate interaction with a helium-rich CSM with a mass of $M_{\rm CSM} \approx 0.1\,M_{\odot}$ and a radius $R_{\rm CSM} \approx 170\,R_{\odot}$ \citep{Yao2020}, which was confirmed by updated modeling in \cite{piro2021}. Both of these are similar to the CSM produced from late stage mass transfer in models from \cite{Wu2022}, motivating a more detailed comparison. These previous works also showed that the light curve is explained by two components, first a shock cooling powered fast rise, followed by a $\sim 0.017 M_{\odot}$ nickel decay-powered peak. Given that we find the HE SCE can also be important, we want to explore if this could be contributing for SN2019dge.

\begin{figure}
\includegraphics[width=0.48\textwidth]{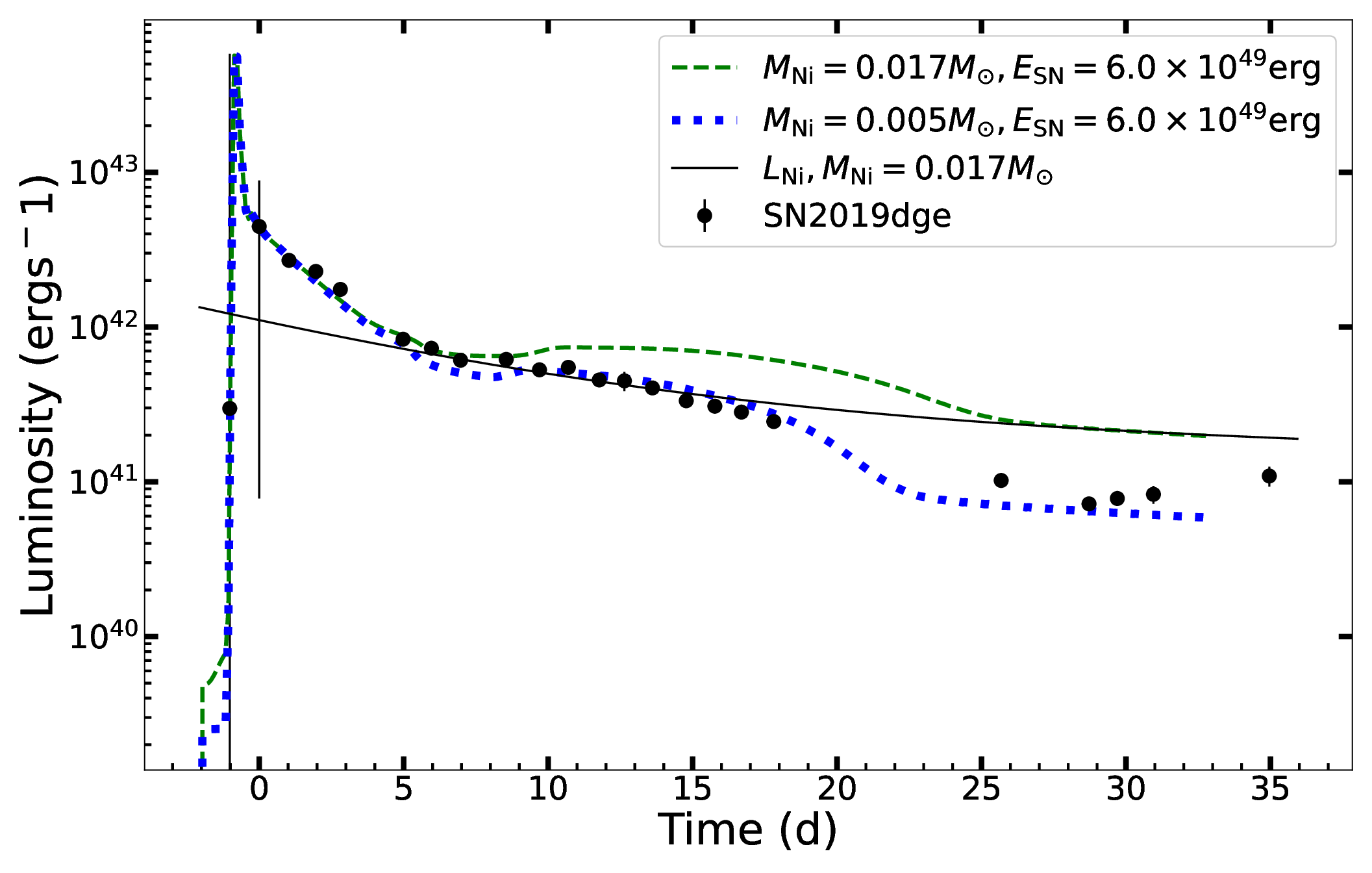}
\caption{Bolometric luminosity of SN2019dge compared to two $M_{\rm He\ Core} = 2.75 M_{\odot}$ models with $M_{\rm Ni} = 0.017 M_{\odot}$ (green dashed) and $0.005 M_{\odot}$ (blue dotted), both exploded with $E_{\rm SN} = 6\times 10^{49}\,{\rm erg}$. Our models favor a lower nickel mass and an important contribution from SCE of the HE to match SN2019dge.}
\label{fig:dge}
\end{figure}

Figure~\ref{fig:dge} compares the bolometric light curve of SN2019dge to the $M_{\rm He\ Core} = 2.75 M_{\odot}$ model, exploded with $E_{\rm SN}=6\times10^{49}\,{\rm erg}$ as is described in Section~\ref{sec:methods} with two different $M_{\rm Ni}$ masses of $0.005$ and $0.017\,M_{\odot}$ as denoted. The black solid line in Figure~\ref{fig:dge} represents the instantaneous heating curve for $M_{\rm Ni} = 0.017\,M_{\odot}$, not accounting for gamma-ray leakage. Given the low energy light curves in Figure~\ref{fig:lightcurves1}, it is fair to assume that gamma-ray leakage has not yet taken over at these early times. Both models have a similar evolution over the first $\approx5\,{\rm days}$ when the luminosity is dominated by CSM SCE. At $\approx10\,{\rm days}$, the observed luminosity indeed matches the nickel heating $M_{\rm Ni} = 0.017\,M_{\odot}$, but when this nickel mass is included in our model (green dashed curve), our simulation predicts too high of a luminosity in comparison to the observations. This is due to the additional contribution from HE SCE. Furthermore, the tail is also too bright compared to what was observed.  If we instead try to match the decay tail, we see a better fit with $M_{\rm Ni} = 0.005\,M_{\odot}$ (blue dotted curve). At this small nickel mass, less energy is required for HE SCE to dominate, so even at $E_{\rm SN} = 6 \times 10^{49}\,$erg there is a luminosity excess that offsets the peak from the corresponding heating curve.

The lower nickel mass model agrees nicely with the light curve of SN2019dge, suggesting that a helium core star that has undergone extreme late-stage mass loss as described in \cite{Wu2022} may be a good candidate for the progenitor of this event. The combination of the especially low explosion energy and nickel mass we infer is qualitatively consistent with the correlations seen for most core-collapse SNe \citep[e.g.,  Figure 3 in][or Figure 8 in \citealp{Lyman2016}]{Kushnir2015}. This is also consistent with the emerging picture of how the core-collapse explosion energy relates to the stellar structure \citep[][and references therein]{Burrows2024}.

\section{Conclusion}\label{sec:conclusion}

We have presented an analytic and numerical investigation of the light curves resulting from  helium-core stars that underwent extreme late-stage mass loss in the years to months before core-collapse \citep{Wu2022}. The structure of these stars at the onset of collapse is similar to that inferred for some observed USSNe and SNe~Ibn, suggesting that this extreme late-stage mass loss mechanism could be a possible evolutionary channel for the progenitors of these events. By applying the analytic framework for shock cooling emission (SCE) discussed in Section~\ref{sec:analytic} to these models, we found that it is possible to observe multiple phases of SCE in a single event, creating three distinct light curve phases: a bright and fast SCE from the circumstellar material (CSM), a longer-lived, dimmer SCE from the HE, and a nickel-powered tail. By numerically generating light curves for these models at varying energies in \texttt{SNEC}, we find that in all models the extended, low density CSM results in a fast, bright initial spike in luminosity as is observed in USSNe and SNe~Ibn. Subsequently, for a given combination of CSM mass, ejecta mass, nickel mass, and explosion energy, it is possible for the HE SCE to outshine the nickel luminosity and also be visible in the light curve as an extended, excess luminosity on timescales of a few to tens of days. 

The visibility of the luminosity excess is sensitive to the HE radius, and we therefore chose to consider this along with the total nickel mass as the driving parameters that set the critical energy above which shock cooling of the helium envelope will dominate over nickel decay. As nickel mass decreases, less energy is required for the light curve to become dominated by HE SCE. The HE SCE feature may therefore be most prominent in USSNe, which tend to have characteristically low nickel masses. 

Finally, we made qualitative comparisons of our models to two rapidly evolving events, SN2019kbj (SN~Ibn) and SN2019dge (USSN). We find promising agreement between our models and the bolometric light curve of SN2019dge by using a lower nickel mass than what has been suggested by previous modeling, which had attempted to match the instantaneous heating curve to the nickel peak at around 14 days. The low nickel mass we infer may actually be naturally expected given the low explosion energy and small overall ejecta mass, as we discuss at the end of Section~\ref{sec:UUSNe}.

For SN2019kbj, we find less agreement between our models and the inferred bolometric light curve, likely due to the ejecta mass of our fiducial model being too low to recreate the long, bright plateau. It does, however, qualitatively capture the general shape of the light curve and supports the possibility of the event originating from a stripped helium star progenitor that has comparable CSM and HE masses. 


The similarities between these events and the models explored in this study support the idea that extreme mass loss in the last years of a helium star's lifetime may explain the characteristics of some USSNe and SNe~Ibn. In modeling observations of other USSNe, it will be important to consider the possibility of some enhancement to the light curve due to HE SCE. This could offset the nickel peak from the heating curve and would result in overestimating the nickel mass when HE SCE is not taken into account.

\acknowledgments

A.H. acknowledges support from the USC-Carnegie fellowship. S.C.W. acknowledges support by the National Science Foundation Graduate Research Fellowship under Grant No. DGE‐1745301. J.F. gratefully acknowledges support through NSF through grant AST-2205974.

\end{document}